\documentclass[10pt]{article}
\usepackage{epsfig,graphicx}
\usepackage{times,ch2005}

\def\beq{\begin{equation}}
\def\eeq#1{\label{#1}\end{equation}}
\def\eeqn{\end{equation}}

\def\beqa{\begin{eqnarray}}
\def\eeqa#1{\label{#1}\end{eqnarray}}
\def\eeqan{\end{eqnarray}}

\def\Dslash{\not{\hbox{\kern-4pt $D$}}}
\def\dslash{\not{\hbox{\kern-2pt $\del$}}}

\def\ee         {\ensuremath{e^-e^-}\xspace}

\newcommand{\tev}{\ensuremath{\mathrm{\,Te\kern -0.1em V}}\xspace}
\newcommand{\gev}{\ensuremath{\mathrm{\,Ge\kern -0.1em V}}\xspace}
\newcommand{\mev}{\ensuremath{\mathrm{\,Me\kern -0.1em V}}\xspace}
\newcommand{\kev}{\ensuremath{\mathrm{\,ke\kern -0.1em V}}\xspace}
\newcommand{\ev}{\ensuremath{\mathrm{\,e\kern -0.1em V}}\xspace}
\newcommand{\gevc}{\ensuremath{{\mathrm{\,Ge\kern -0.1em V\!/}c}}\xspace}
\newcommand{\mevc}{\ensuremath{{\mathrm{\,Me\kern -0.1em V\!/}c}}\xspace}
\newcommand{\gevcc}{\ensuremath{{\mathrm{\,Ge\kern -0.1em V\!/}c^2}}\xspace}
\newcommand{\mevcc}{\ensuremath{{\mathrm{\,Me\kern -0.1em V\!/}c^2}}\xspace}

\def\mus  {\ensuremath{\rm \,\mus}\xspace}

\def\mus        {\ensuremath{\,\mu{\rm s}}\xspace}    
\def\be{\begin{equation}}
\def\ee{\end{equation}}
\def\lsim{\lower 2pt \hbox{$\, \buildrel {\scriptstyle <}\over
         {\scriptstyle \sim}\,$}}
\def\gsim{\lower 2pt \hbox{$\, \buildrel {\scriptstyle >}\over
         {\scriptstyle \sim}\,$}}

\begin{document}


\Title{1-100 GeV emission from millisecond pulsars}
\bigskip


%
\label{HardingStart}

%
\author{Alice K. Harding$^1$\index{Harding, A. K.}, Vladimir V. Usov$^{3}$
\index{Usov, V.V.} and Alex Muslimov$^{1,2}$\index{Muslimov, A.}}

%
\address{$^1$NASA Goddard Space Flight Center\\
Greenbelt, MD 20771, USA \\ 
$^2$Universities Space Research Association\\
$^3$Weizmann Institute of Science\\
Rehovat 76100, Israel}

\makeauthor\abstracts{
A number of rotation-powered millisecond pulsars are powerful sources of X-ray emission. We present predictions for the spectral characteristics of these sources at gamma-ray energies, using a model for acceleration and pair cascades on open magnetic field lines above the polar caps. Since these pulsars have low surface magnetic fields, the majority do not produce sufficient pairs to completely screen the accelerating electric field allowing particle acceleration to high altitude. The resulting emission above 1 GeV comes from curvature radiation by primary electrons with radiation-reaction-limited Lorentz factors. The spectra are very hard power-laws with exponential cutoffs between 1 and 50 GeV, and the spectral power peaks near the cutoff energy. Millisecond pulsars are thus ideal targets for air-Cherenkov detectors that are able to reach energy thresholds below 50 GeV.}

\section{Introduction}

There are more than 100 radio pulsars now known with periods between 1 and 30 milliseconds.  They are believed 
to have been spun-up by accretion torques due to mass flow from a binary companion \cite{Alpar82} and
have surface magnetic fields, in the range of $10^8 - 10^{10}$ G, much lower than those of the normal
pulsar population.  Most are found in binary systems with low mass or compact (White Dwarf or neutron star) 
companions.  Although they are found throughout the Galaxy, globular clusters seem to be particularly good 
breeding grounds for millisecond pulsars (MSPs) and a number of clusters contain as many as 15 (in 47 Tuc) 
or 21 (in Ter 5) \cite{Ransom05}.  
Many MSPs are also seen at high energy wavelengths.  About 30 MSPs are detected as X-ray point
sources and pulsed emission has been detected in 7 of these \cite{ba02, Kaspi05}.   
Pulsed $\gamma$-ray emission has been detected from one MSP, PSR J0218+4232, by EGRET on the Compton Gamma-Ray
Observatory \cite{Kuiper00}.  Thus, MSPs must be powerful particle accelerators and despite their low
surface magnetic fields, their spin-down luminosities are comparable to those of young pulsars.

Polar cap particle acceleration theory \cite{brd00, lsm00, HMZ02, HVM05} predicts that MSPs 
should accelerate particles to high Lorentz factors ($\sim 10^7$).  These high energies, combined with the
small radius of curvature of the open field lines, enables curvature radiation emitted by these
particles to reach 10 - 100 GeV, depending on the pulsar parameters.  Most should be easily detectable by 
GLAST and some may be detectable by Air Cherenkov Telescopes (ACTs).  We will discuss how the predicted spectrum
and flux depend on the period, magnetic field, inclination and equation of state of the pulsar and the prospects for
detection of the emission by various instruments.

\section{Curvature Radiation from Accelerated Particles}

MSPs have the lowest magnetic fields of any radio pulsars and such low fields have two important consequences
for their high-energy emission.  They produce relatively few electron-positron 
pairs by one-photon pair production, a process that requires fields nearer the critical value of 
$B_{cr} = 4.4 \times 10^{13}$ G.  As a result, the pair cascades in most MSP magnetospheres will not screen the
electric field as they do in younger pulsars.  Also, the spectral cutoff due to pair production attenuation,
which depends inversely on field strength, will occur at much higher energy.  Therefore, particle energies
will be limited by curvature radiation reaction to a value \cite{lsm00, HMZ02, HVM05}
\be
\gamma _{_{CRR}}  = \left( {\frac{3}{2}\frac{{E_{||} \rho _c^2 }}{e}} \right)^{1/4},
\ee
where $E_{||}$ is the parallel electric field and $\rho _c$ is the field line radius of curvature.  The 
parallel electric field, including the contribution from inertial frame-dragging, is \cite{mt92, HVM05}
\be \label{E||}
E_{_{||} }^{}  \cong 5 \times 10^5 B_8 P_{ms}^{ - 2} \left[ {\kappa _{0.15}^{} \frac{{\cos \chi }}{{\eta ^4 }} + \frac{{\theta _0 }}{4}\eta ^{ - 1/2} \sin \chi \cos \phi } \right],
\ee
where $B_8 \equiv B/10^8$ G is surface magnetic field strength, $P_{ms} \equiv P/1$ ms is pulsar period, 
$\chi$ is the inclination angle, $\phi$ is magnetic azimuth angle, $\theta_0 = (2\pi R/c P)^{1/2}$ is the
polar cap half-angle, 
$\eta = r/R$ is the radial coordinate and $\kappa _{0.15} \equiv
0.15 I_{45}/R_6^3$, $I_{45} \equiv I/10^{45}\,\rm g\,cm^2$ and $R_6 \equiv R/10^6$ cm are the moment or inertia and radius
of the neutron star. In this case, the curvature radiation peak energy, $\varepsilon _{peak}^{CR} = (4/3)
\varepsilon _{CR}$ will be
\be
\varepsilon _{peak}^{CR}  = 2\frac{{\lambda _C \gamma _{_{CRR} }^3 }}{{\rho _c }} \approx 10\,{\rm{GeV}}\,B_8^{3/4} \left\{\begin{array}{*{20}c}
   {P_{ms}^{ - 5/4} \kappa _{0.15}^{3/4} \eta ^{ - 11/4} \begin{array}{*{20}c}
   {} & {}  \\
\end{array}\chi  = 0}  \\
   {P_{ms}^{ - 3/8} \eta ^{ - 1/8} \begin{array}{*{20}c}
   {} & {} & {}  \\
\end{array}\chi  = 
{\pi  \mathord{\left/
 {\vphantom {\pi  2}} \right.
 \kern-\nulldelimiterspace} 2}}  \\
\end{array} \right.
\ee
where $\lambda _C$ is the electron Compton wavelength.  Since the polar cap half-angle $\theta_0$ is large
for MSPs, the radius of curvature of open field lines, $\rho_c \simeq r/\theta_0$, is relatively small.
Thus the CR spectrum of MSPs can extended beyond 10 GeV, but since the pair creation will attenuate the emission
above $\varepsilon_{\rm esc } \sim 10^6~P_{\rm ms}^{1/2}B_8^{-1}$, 
where $\varepsilon_{\rm esc }$ is the
photon escape energy in $mc^2$, 
the maximum energy is effectively limited to
100 GeV or below.  

It should be noted that any detected emission above 2 GeV from MSPs cannot be produced in outer gap accelerators \cite{zc03, hir04},
since the quantity $E_{||}^{3/4} \rho _c^{1/2}$ determining the peak energy is smaller.
Therefore, emission detected from MSPs at very high energies must originate from near the polar cap.

\begin{figure}[t] 
\vskip -1.0cm
\begin{center}
\epsfig{file=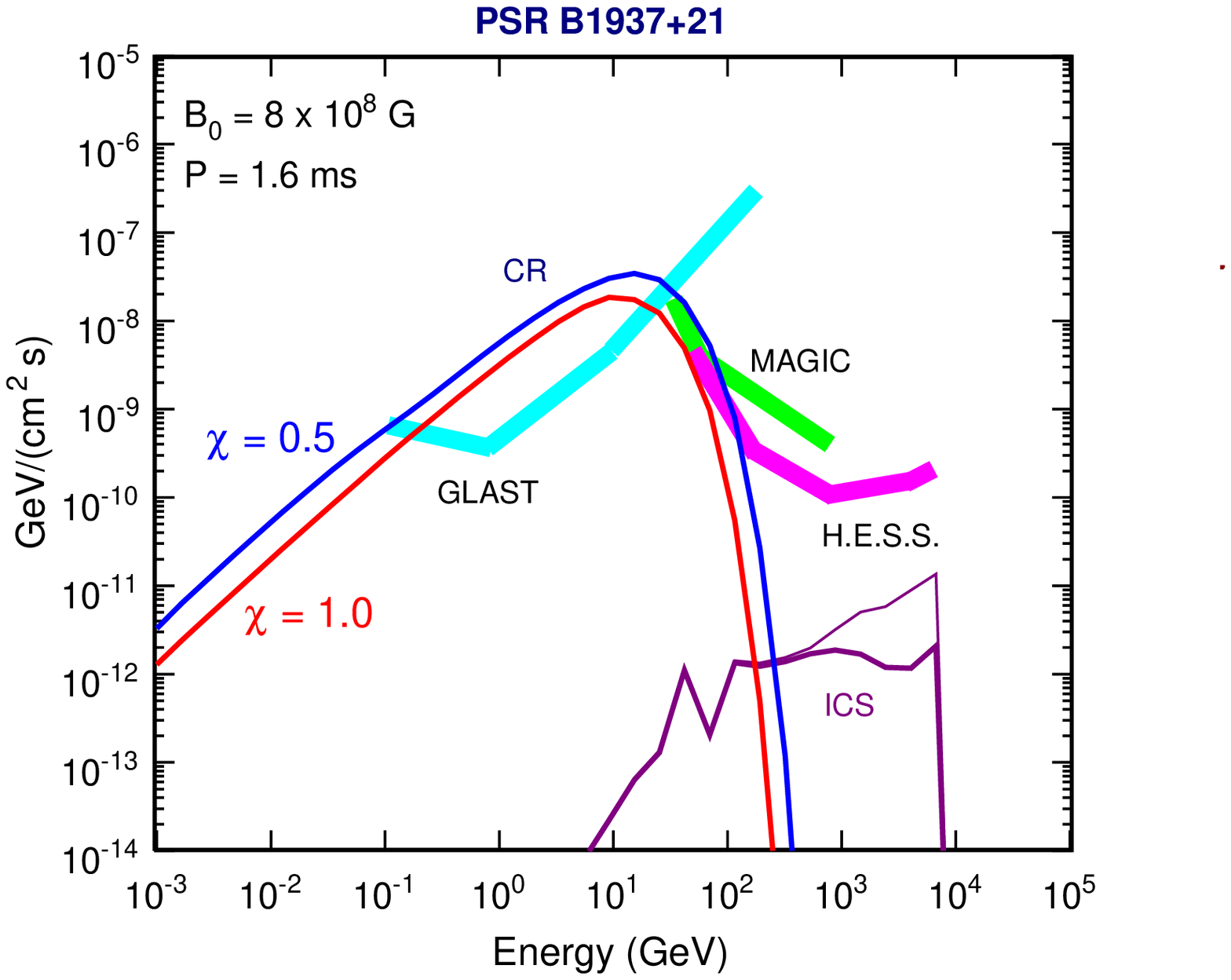,width=13cm}
\caption{Predicted curvature radiation (CR) and unattenuated (light line) and attenuated (dark line) 
inverse Compton (ICS) spectra of PSR B1937+21, for two different inclination angles $\chi$ in radians and
FP equation of state (see Fig. 2).}
\label{fig:Harding-fig1}
\end{center}
\end{figure}

\begin{figure}[t]
\vskip -0.7cm
\begin{center}
\epsfig{file=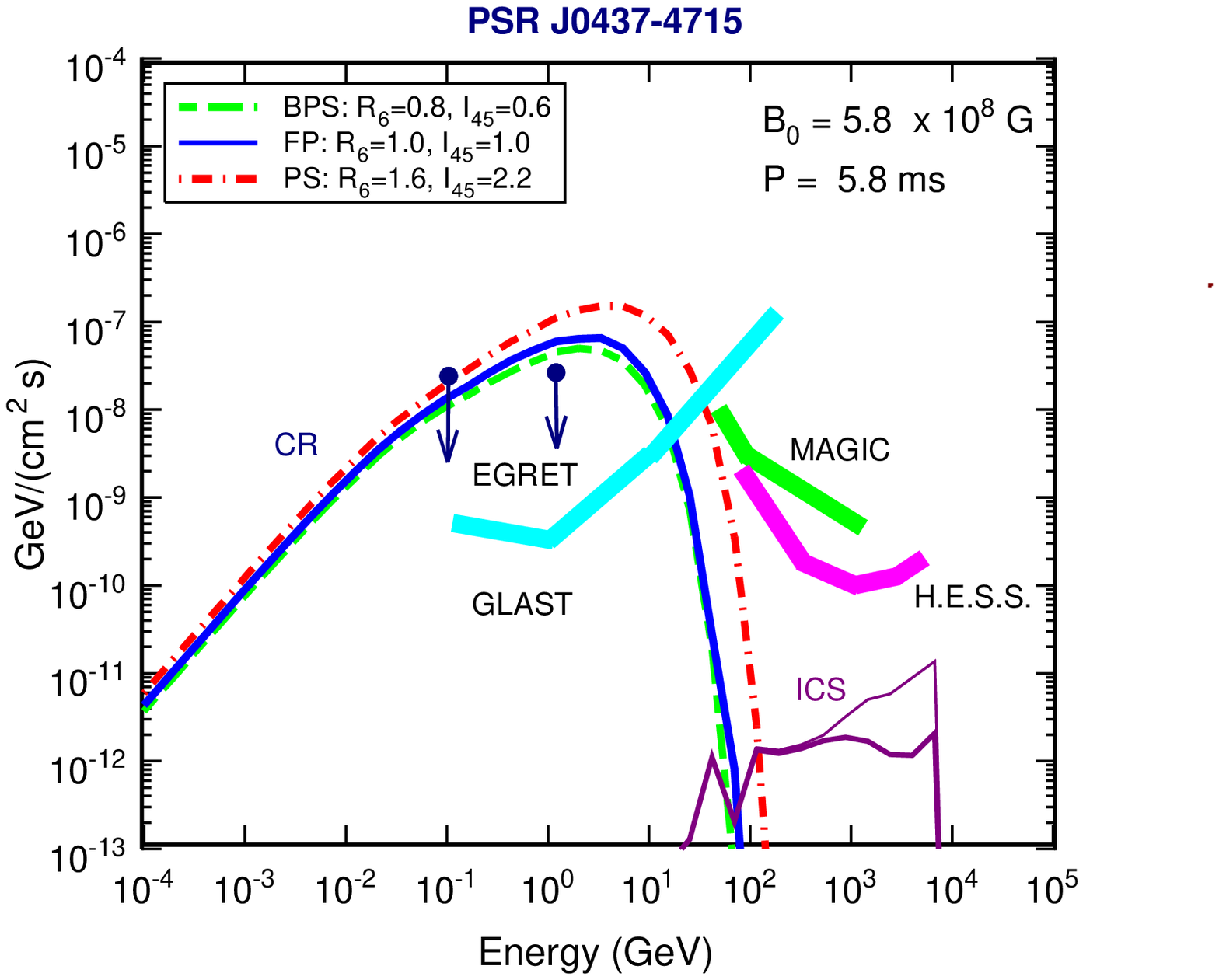,width=13cm}
\caption{Predicted curvature radiation (CR) and unattenuated (light line) and attenuated (dark line) 
inverse Compton (ICS) spectra of PSR J0437-4715, for soft (BPS), medium (FP) and hard (PS) neutron star
equations of state (see \cite{HMZ02}) and inclination angle $\chi = 0.1$. EGRET upper limits are from
\cite{vdj05}.}
\label{fig:Harding-fig2}
\end{center}
\end{figure}

\section{Predicted Spectrum}

The predicted curvature radiation spectrum of electrons with radiation-reaction limited Lorentz factors
is very hard, with photon power law index $-2/3$ and exponential cutoff $\exp{(-\varepsilon/\varepsilon _{CR})}$.
Figure 1 shows the calculated spectrum of curvature radiation for PSR B1937+21, which is the fastest MSP with
period of 1.6 ms.  It has a predicted CR peak energy around 25 GeV, which is not attenuated by pair creation since 
$\varepsilon_{\rm esc } \sim 30$ GeV, so that significant emission extends above 50 GeV.  
The spectrum is shown for two different inclination angles and
since the magnetic axis of this pulsar may be highly inclined, from the fact that both the radio and X-ray profiles
have interpeaks, it may be more difficult to detect with H.E.S.S. or VERITAS.  A component due to inverse Compton scattering (ICS) of thermal X-rays from the neutron star surface by accelerated electrons is also expected and will
peak at energies of 1-10 TeV.  However, the ICS emission will be strongly attenuated by magnetic pair production
to a level that is far below present detector sensitivities.

Figure 2 shows the calculated spectrum of CR for PSR J0437-4715, illustrating the dependence of the emission 
on the neutron star equation of state (EOS).  The values of $R_6$ and $I_{45}$ for the different EOS determine
the value of the $\kappa$ parameter in the parallel electric field of eqn (\ref{E||}).  
The apparent presence of a core component in the radio profile
suggests a small viewing impact angle to the magnetic pole.  Therefore, if the acceleration model described above is
correct, we should see the full low-altitude CR component.  The predicted spectrum for the stiffer EOS (PS) 
is not consistent with the EGRET upper limits, suggesting that either the EOS is soft or that the model is incorrect. 
In the former case, PSR J0437-4715 will not be detectable by H.E.S.S.

\begin{figure}[t]
\vskip -0.5cm
\begin{center}
\epsfig{file=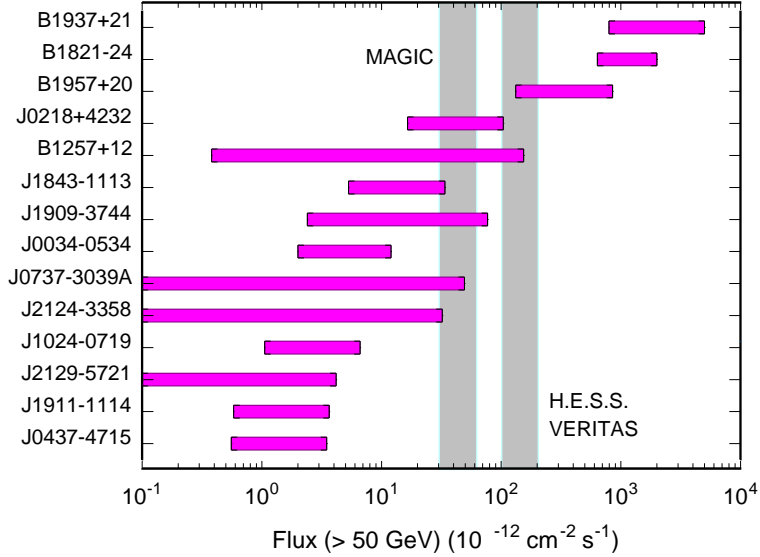,width=11cm}
\caption{Predicted range of integral flux ($> 50$ GeV) from individual millisecond pulsars.  Light gray vertical
bands are integral flux thresholds at $> 50$ GeV for H.E.S.S., VERITAS and MAGIC II \cite{ow05}}
\label{fig:Harding-fig3}
\end{center}
\end{figure}

\section{Predicted Flux From MSPs in the Galactic Plane}

Figure 3 displays a rank-ordered list of MSPs and their predicted integral fluxes $> 50$ GeV, where the flux
range reflects the dependence on inclination angle.  Only a few MSPs, including B1937+21, B1821-24 and B1957+20, 
have a flux as predicted by this model that would be detectable above 50 GeV and all three have pulsed X-ray
emission.  Actually, B1821-24 is the only
known MSP in the globular cluster M28, but at a distance of 4.9 kpc may be individually detectable.
A few more MSPs would be detectable by ACTs such as MAGIC that may reach energy thresholds as low as 30 GeV and that 
may have lower flux sensitivity above 50 GeV.  J0218+4232 is the only MSP detected above 100 MeV by EGRET 
\cite{Kuiper00} and may be detectable by ACTs depending on the viewing angle.  
J0737-3039A is the MSP with $P = 23$ ms in the double pulsar
binary system \cite{Lyne04} and whose radio emission is eclipsed by the magnetosphere of its 2.8 s pulsar companion.

\begin{figure}[t]
\vskip -0.5cm
\begin{center}
\epsfig{file=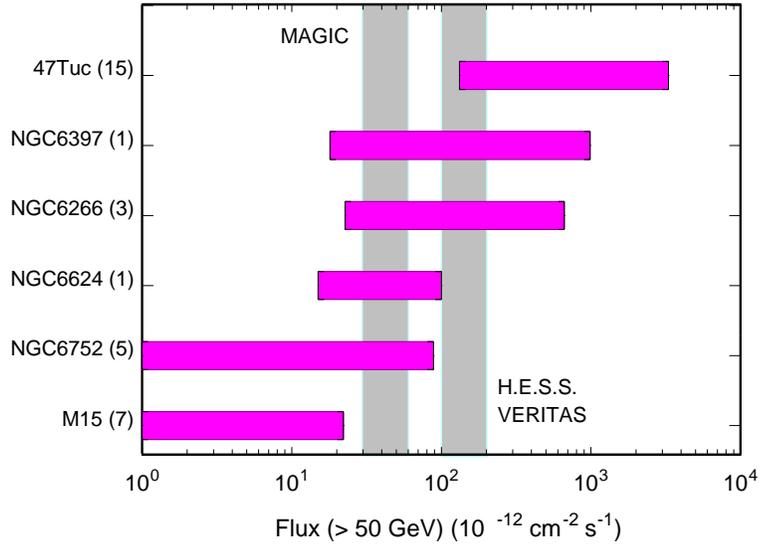,width=11cm}
\caption{Predicted integral flux ($> 50$ GeV) from globular clusters containing millisecond pulsars.  The
number of known radio pulsars are shown in parentheses for each cluster.}
\label{fig:Harding-fig4}
\end{center}
\end{figure}

\section{Predicted Flux From Globular Clusters}

Although globular clusters are generally more distant than many individual Galactic MSPs, the combined flux
from multiple MSPs in these clusters may be detectable by ACTs.  Figure 4 shows several globular clusters 
containing one or more MSPs, rank-ordered by their predicted combined flux $> 50$ GeV from the known radio pulsars
in the cluster.  In particular 47 Tuc, containing
fifteen known radio MSPs and possibly many more undiscovered MSPs, may be detectable by ACTs.  The clusters
NGC6397 and NGC6624 each contain only one known MSP, J1740-5340 and J1823-3021A respectively, 
that individually may produce
a detectable flux.  However, there is a greater degree of uncertainty in predicting $\gamma$-ray flux from MSPs
in globular clusters since the measured period derivatives have model-dependent contributions from acceleration in
the cluster potential \cite{Rob95}.

\section{Conclusions}

Pulsed emission from several millisecond pulsars may be detectable by ACTs at energy greater than 50 GeV, 
if emission from a low-altitude accelerator is visible.  In particular, 
telescopes that are able to reach thresholds below 50 GeV
should be able to detect all the X-ray bright MSPs.  For large viewing angle to the magnetic axis, the low-altitude
emission which reaches the highest energies will not be visible.  In this case, even if emission from an outer gap accelerator is visible, no emission above 10 GeV is expected.  Several globular clusters, including 47 Tuc, 
NGC6397 and NGC6266, may also be detectable by ACTs.  Detection or limits for individual pulsars believed to have 
small viewing angles to the magnetic pole, such as B1937+21 and J0437-4715 will give direct constraints on the electrodynamics of particle acceleration.

%
\label{HardingEnd}
 
\end{document}